\pdfoutput=1 
\documentclass[11pt,a4paper]{article}

\usepackage{booktabs} 
\usepackage[utf8]{inputenc}
\usepackage[T1]{fontenc}
\usepackage{amsmath,amssymb,amsfonts,amsthm} 
\usepackage{graphicx}                         
\usepackage[section]{placeins}                
\usepackage{xcolor}
\usepackage{geometry}                         
\usepackage{authblk} 

\title{\textbf{From Zero to Mega-Gauss Fields: Comprehensive Magnetophotonic Spectroscopy of Graphene Dirac Cones}}

\author{
  Shojiro Takeyama\thanks{takeyama@issp.u-tokyo.ac.jp} \quad and \quad
  Hiroaki Saito\thanks{Present address: Nissan Motor Co., Ltd., 1-1, Takashima 1-chome, Nishi-ku, Yokohama-shi, Kanagawa 220-8686, Japan} \\
  \vspace{2mm} 
  \normalsize \textit{The International MegaGauss Science Laboratory, The Institute for Solid State Physics,} \\
  \normalsize \textit{The University of Tokyo, 5-1-5 Kashiwanoha, Kashiwa 277-8581, Japan}
}
\usepackage{hyperref}
\usepackage[all]{hypcap} 

\hypersetup{
    colorlinks=true,
    linkcolor=blue,
    citecolor=blue,
    urlcolor=blue
}
\usepackage{upgreek}
\usepackage{amsmath}
\usepackage{amssymb}
\usepackage{bm}
\usepackage{siunitx}
\usepackage{tabularx}
\begin{document}

\maketitle


\begin{abstract}
We investigate the infrared magneto-optical response of n-doped epitaxial graphene on 4H-SiC in ultrahigh magnetic fields up to 560~T, utilizing single-turn coil and electromagnetic flux compression techniques. The measured absorption spectra are anomalously broad, strongly deviating from conventional cyclotron resonance. Angle-resolved photoemission spectroscopy (ARPES) reveals a distorted Dirac dispersion featuring a ``camel-back" structure with an energy gap of $E_g \sim 0.2$~eV. Using the band parameters extracted via a generalized bilayer graphene model, we construct a Landau level (LL) fan chart that dictates a critical level crossing between the $N=0^+$ and $N=0^-$ states near 160--200~T. At this threshold, the optical transition mechanism undergoes a dramatic shift from an electron-dominated collective mode to a cooperative electron-hole collective excitation. Furthermore, the extreme-field absorption spectra under thulium fiber laser excitation ($\hbar\omega_0 = 0.636$~eV)---culminating in a massive resonance near 400~T with a shoulder at 200~T---are excellently reproduced by a collective Alfv\'en wave model. 
This analysis also evidences a magnetic-field-induced enhancement of the sublattice potential asymmetry parameter (from 0.10~eV to 0.12~eV). This directly signifies that the macroscopic electron-hole band asymmetry is further amplified by the applied magnetic field.
Ultimately, the field-induced energy inversion generates a strongly interacting, fully compensated electron-hole plasma. The resonant excitation of Alfv\'en waves in this regime demonstrates that a pristine tabletop analog to the relativistic electron-positron plasmas found in astrophysical extremes can be elegantly realized within a 2D graphene system.
\end{abstract}


\vspace{1em}
\noindent \textbf{Keywords:} Ultrahigh magnetic field, Magneto-optical spectroscopy, Angle-resolved photoemission spectroscopy, Graphene, Alfv\'en wave, Distorted Dirac dispersion

\clearpage 

\section{Introduction}

The Dirac dispersion relation fundamentally describes the energy-momentum kinematics of massless relativistic particles, such as photons, governing the fundamental laws of the universe. In the realm of solid-state physics, this relativistic physics is elegantly realized in graphene---an ideal two-dimensional (2D) system spanned by a monoatomic honeycomb lattice, where the motion of electrons is strictly governed by the Dirac equation. Graphene thus provides a unique microcosm, often affectionately termed ``the universe on a desktop,'' allowing researchers to explore extreme relativistic phenomena in a tabletop experiment~\cite{Geim2007}. Furthermore, within the broader context of condensed matter physics, graphene serves as the ultimate fundamental building block, offering a platform to investigate the diversity and universality of physical properties across spatial dimensions---spanning from a perfect 2D monolayer to bilayer, few-layer, and ultimately the three-dimensional (3D) bulk graphite.

Following the successful isolation of graphene via mechanical exfoliation by Novoselov \textit{et al.} in the early 2000s, an explosive volume of research was initiated worldwide. Extensive transport and spectroscopic studies have continuously unveiled the profound physical properties stemming from the Dirac cone. These milestones are beautifully summarized in comprehensive experimental and theoretical reviews \cite{Novoselov2011,CastroNeto2009,Orlita2010,Ando2007,Ando2008}. Among its most striking features are phenomena rooted in quantum electrodynamics (QED): the chiral symmetry of electrons and holes (analogous to particles and antiparticles), the Klein paradox leading to perfectly enhanced tunneling (where the transmission probability becomes exactly 1), the absence of backscattering resulting in a constant Fermi velocity, the half-integer quantum Hall effect, the universal infrared limit of optical absorption ($\pi\alpha \approx 2.3\%$), and strict Pauli blocking. Remarkably, all these phenomena are deeply intertwined with the magnetic flux quantum and the fine-structure constant ($\alpha$), explicitly manifesting the effective ``speed of light'' in macroscopic electrical and optical observations.

To investigate whether the robust $\sqrt{B}$ scaling of Dirac fermions persists up to the extreme quantum limit, we previously conducted infrared magneto-absorption measurements in ultrahigh magnetic fields up to 560~T \cite{Nakamura2020}. While the observed deviations from the ideal $\sqrt{B}$ dependence were initially treated as phenomenological higher-order corrections within the $\mathbf{k}\cdot\mathbf{p}$ perturbation scheme \cite{AndoSuzuura2017}, a careful reassessment of this approach raises a fundamental question. The necessity of invoking such higher-order terms to explain the anomalies implies that the perfectly linear Dirac model is inherently insufficient. Rather than viewing these deviations as mere perturbative effects, they strongly indicate that the system is governed by an intrinsically ``distorted'' Dirac dispersion from the outset, characterized by relativistic electron-hole symmetry breaking.

This realization naturally provokes a more fundamental question: to what extent is the ideal linear Dirac dispersion actually maintained in real systems? It is highly plausible that the simple introduction of carriers via electron doping, or even the optical excitation of electrons into the conduction band, is sufficient to trigger spontaneous electron-hole symmetry breaking, thereby inherently transforming the system into a distorted Dirac dispersion.

In this paper, we systematically revisit our raw experimental data and incorporate an analysis of angle-resolved photoemission spectroscopy (ARPES) measurements to explicitly demonstrate that highly doped graphene exhibits a fundamentally distorted Dirac dispersion. 
Based on this established band structure, we calculate the generalized Landau level (LL) fan charts. 
We reveal that the magneto-absorption spectra observed in doped samples during the single-turn coil (STC) measurements cannot be ascribed to conventional single-particle CR. Rather, they represent an electron-dominated collective plasma mode (e.g., helicon waves) that acts as a precursor to the fully cooperative electron-hole plasma phase realized at extreme magnetic fields. Furthermore, we establish that the anomalously broad absorption spectra observed in the extreme magnetic field regime of the electro-magnetic flux compression (EMFC) (up to 500~T) are not single-particle transitions, but are manifestations of Alfv\'en-wave-like collective modes propagating in a highly dense electron-hole plasma.

\section{Experimental Methods}
\label{sec:experimental}

Epitaxial graphene was synthesized by the thermal decomposition of a Si-terminated 4H-SiC substrate \cite{Hibino2008,Hibino2010,Hibino2012}. Micro-Raman mapping and atomic force microscopy (AFM) images confirmed that 70--80\% of the surface area was covered by monolayer graphene (for details, see the Supplemental Material of Ref.~\cite{Nakamura2020}). This type of graphene is naturally electron-doped due to charge transfer from the SiC substrate \cite{Zhou2007}. The measurements were conducted on two types of doped samples at room temperature. Electrical properties, including the carrier density and mobility, were characterized using the van der Pauw method, with the estimated parameters summarized in Table~\ref{tab:carrier_properties}. For the magneto-optical measurements, the graphene samples were shaped into a circular form with a diameter of 2.5~mm, which were core-cut from a larger $10 \times 10\text{ mm}^2$ sheet, while the laser spot diameter was maintained within 1.6~mm (See the Supplementary Material for details~\cite{SM}) .

\begin{table}[htbp]
\centering
\caption{\label{tab:carrier_properties} \footnotesize Carrier density $n$, mobility $\mu$, and Fermi energy $E_F$ of each sample estimated by the van der Pauw method.}
\begin{tabular}{lccc}
\hline \hline 
\textbf{Sample} & \textbf{\begin{tabular}[c]{@{}c@{}}$n$\\ ($\times 10^{11}$ cm$^{-2}$)\end{tabular}} & \textbf{\begin{tabular}[c]{@{}c@{}}$\mu$\\ (cm$^2$/Vs)\end{tabular}} & \textbf{\begin{tabular}[c]{@{}c@{}}$E_F$\\ (meV)\end{tabular}} \\
\hline 
Sample A & 3 & 3000 & 50 \\
Sample B & 7 & 1000 & 90 \\
\hline \hline
\end{tabular}
\end{table}

Ultrastrong magnetic fields in the ranges of 0--200~T and 0--600~T were generated using the STC and EMFC methods, respectively. Detailed descriptions of these megagauss generation techniques, along with the infrared magneto-optical measurement configurations, can be found in Ref.~\cite{Takeyama2026} (see Fig.~48 therein). For the STC measurements, linearly polarized infrared gas lasers---specifically, CO$_2$ ($\lambda = 9\text{--}11\,\mu\text{m}$), CO ($\lambda = 5.2\text{--}5.7\,\mu\text{m}$), and He--Ne ($\lambda = 3.39\,\mu\text{m}$) lasers---were utilized as optical sources. In the EMFC experiments, a thulium fiber laser ($\lambda = 1.95\,\mu\text{m}$) was employed. 

During the EMFC infrared transmission measurements, special care must be taken to mitigate stray infrared light emitted from the imploding liner, which otherwise severely disrupts optical detection as the magnetic field approaches its maximum. The specific techniques developed to overcome this challenge are detailed in the Supplementary Material~\cite{SM}. The optical transmission signals were recorded using a photovoltaic HgCdTe detector responsive over the $\lambda = 2\text{--}12\,\mu\text{m}$ wavelength range.

ARPES measurements were performed using the apparatus at the Institute for Solid State Physics (ISSP), The University of Tokyo. A helium discharge lamp targeting the He~II line ($h\nu = 40.8\,\text{eV}$) was utilized as the excitation source, with an energy resolution maintained at $27\,\text{meV}$. The typical X-ray beam spot area on the sample was $2 \times 2\,\text{mm}^2$. Prior to the photoemission measurements, the samples were annealed at $500\,^\circ\text{C}$ for 30~minutes under ultra-high vacuum conditions to desorb any surface-adsorbed molecules and ensure an atomically clean surface.

\section{Results and Discussion}
\label{sec:ResultDiscus}
\unskip
\subsection{Infrared Magneto-Absorption in STC Measurements}

\begin{figure}[htbp]
\centering 
\includegraphics[width=0.6\columnwidth]{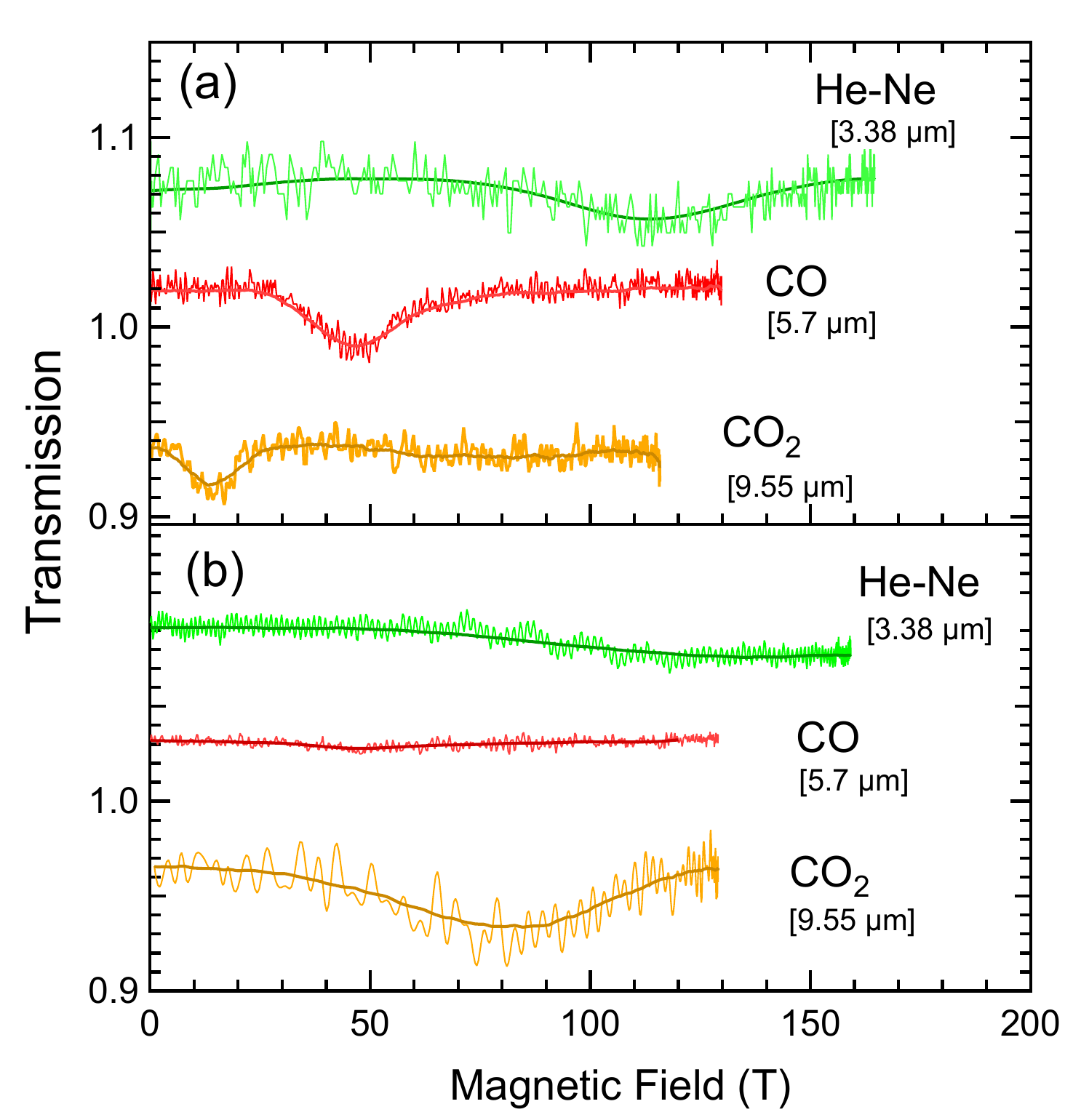}
\caption{\label{fig:waveformSTC}
\footnotesize Magneto-transmission spectra in magnetic fields measured using He-Ne, CO, and CO$_2$ lasers with photon energies of $\hbar\omega_{0} = 0.367\,\text{eV}$, $\hbar\omega_{0} = 0.218\,\text{eV}$, and $\hbar\omega_{0} = 0.130\,\text{eV}$, respectively, for (a) sample A, and (b) sample B. (Note: For Sample A [Fig.~\ref{fig:waveformSTC}(a)], the raw experimental data are identical to those presented in Fig. 1 of Ref.~\cite{Nakamura2020}.)} 
\end{figure}

Figure~\ref{fig:waveformSTC} displays the infrared magneto-transmission spectra obtained from the STC experiments. 
Here, the results for both Sample A and Sample B are plotted for three different incident light wavelengths: $\rm{CO_2}$ ($9.55~\mu\text{m}$), $\rm{CO}$ ($5.7~\mu\text{m}$), and $\rm{He\text{-}Ne}$ ($3.38~\mu\text{m}$). 
These weak absorption profiles stand in stark contrast to standard two-dimensional electron gases (2DEGs) hosted in semiconductor heterostructures, such as InAs/AlSb single quantum wells; at similar carrier density regimes ($5 \times 10^{11}$--$12 \times 10^{11}~\text{cm}^{-2}$), they routinely demonstrate robust CR features with transmission depths as large as 20--30\%~\cite{Arimoto2003}.

For Sample A [Fig.~\ref{fig:waveformSTC}(a)], the resonance dips observed for the lower-energy $\rm{CO_2}$ and $\rm{CO}$ lasers can be regarded as conventional single-particle CR. This assignment is justified both by the magnitude of their absorption intensities and by their corresponding resonance fields ($B_{\rm{rs}}$), which are rather consistent with the values predicted by the characteristic $\sqrt{B}$ dependence of Dirac fermions. However, their absorption spectra are surprisingly broad and exhibit a tail extending toward the lower-magnetic-field side, a lineshape that is strictly opposite to what is expected for graphene (as described by Eq.~S1 in S2, Supplementary Material~\cite{SM}). Furthermore, for the higher-energy $\rm{He\text{-}Ne}$ photon probe, the spectral curve becomes even more anomalously broad, and the absorption intensity is substantially weakened.

Furthermore, in the highly doped Sample B [Fig.~\ref{fig:waveformSTC}(b)], none of the observed spectra resemble typical CR. Attempts to interpret these features based on the standard $\sqrt{B}$ scaling yield entirely contradictory results. 
These systematically suppressed and heavily distorted profiles strongly suggest that the observed magneto-optical features cannot be ascribed to conventional single-particle CR transitions, but rather signify the onset of a distinct collective absorption mechanism.

In our previous study~\cite{Nakamura2020}, the magneto-absorption spectra were entirely analyzed within the framework of conventional CR, assuming the characteristic $\sqrt{B}$ scaling dictated by an ideal linear Dirac dispersion. 
However, to account for the anomalous double-peak structures observed in the extreme magnetic field range of 200--500~T, we incorporated a linear momentum term based on the theory by Ando and Suzuura~\cite{AndoSuzuura2017}. 
Consequently, we concluded that ``the observed CR splitting is interpreted as a consequence of the electron-hole asymmetry of the energy band dispersion caused by the higher-order term in the $\mathbf{k}\cdot\mathbf{p}$ scheme'' [Eqs.~(2) and (3) in Ref.~\cite{Nakamura2020}].
A careful reassessment of this logic, however, raises a fundamental question. The necessity of invoking electron-hole asymmetry and higher-order $\mathbf{k}\cdot\mathbf{p}$ perturbations inherently challenges the starting assumption of a perfectly linear Dirac dispersion. Rather than a mere perturbative correction at extremely high fields, this logical inconsistency strongly implies that the intrinsic band structure of this system is governed by a substantially distorted Dirac dispersion from the very outset.

\subsection{ARPES and Distorted Dirac Dispersion}
\label{sub:arpes}
To experimentally address this fundamental question and directly verify the existence of such an intrinsically distorted band structure, we proceeded to investigate the electronic dispersion via ARPES. Historically, ARPES measurements on monolayer graphene exfoliated via the micromechanical cleavage (``Scotch-tape'') method have been exceedingly difficult, resulting in a limited number of reports. This difficulty primarily arises from an inherent size mismatch: typical exfoliated graphene flakes are on the order of a few micrometers, which is significantly smaller than the beam spot size of conventional ARPES instruments (typically ranging from tens to hundreds of micrometers). 

While specialized techniques such as Nano-ARPES have been utilized to successfully capture the Dirac cone in supported exfoliated flakes \cite{Knox2008}, epitaxial graphene grown on SiC substrates offers a more practical and robust platform by providing large-area, high-quality samples. Consequently, epitaxial monolayer graphene has been extensively investigated using ARPES, yielding pivotal insights into its electronic properties. Representative studies include the detailed observation of quasiparticle dynamics and many-body interactions \cite{Bostwick2007}, the demonstration of a substrate-induced bandgap opening driven by A-B sublattice symmetry breaking \cite{Zhou2007}, and the decoupling of the buffer layer via hydrogen intercalation to restore a quasi-free-standing gapless Dirac dispersion \cite{Riedl2009}. 

\begin{figure}[htbp]
\centering
\includegraphics[width=0.7\columnwidth]{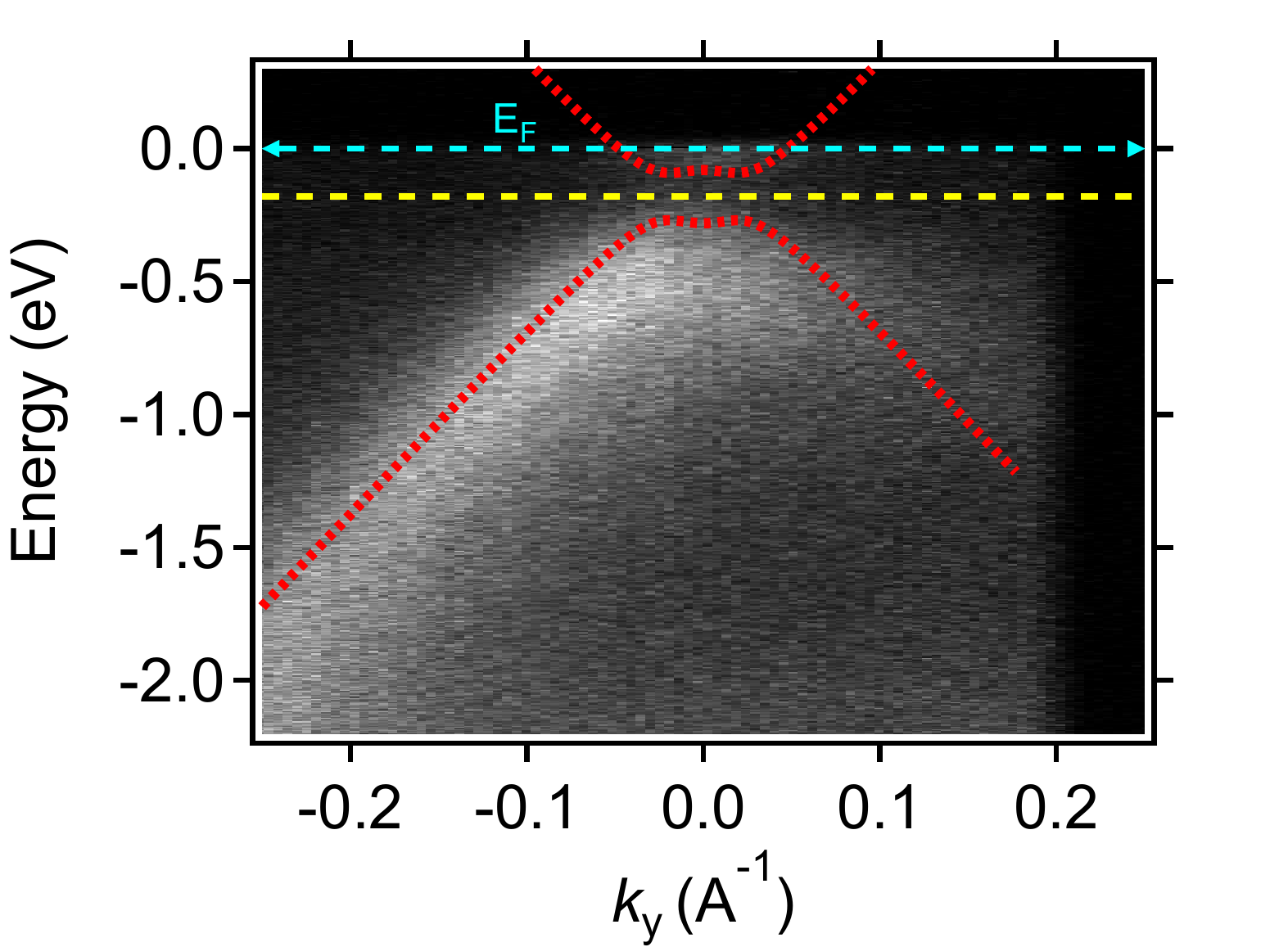}
\caption{\label{fig:ARPES}
\footnotesize Two-dimensional ARPES energy-momentum intensity map captured along the $k_y$ direction across the $K$ ($K'$) high-symmetry points. 
The red solid line denotes the best-fit curve calculated from the model, Eq.~(\ref{eq:deformedDirac}), which cleanly traces the distorted dispersion Dirac band. The center of the energy gap (the yellow dashed line) is positioned approximately 0.18~eV below the Fermi energy ($E_F$), indicated by the sky blue dashed line.}
\end{figure}

Building upon these foundational ARPES studies, our measurements on heavily electron-doped epitaxial graphene reveal details of the  $E-k$ band dispersion.
Figure~\ref{fig:ARPES} displays the electronic band dispersion of the graphene sample B obtained by ARPES. The two-dimensional (2D) energy-momentum intensity map was acquired by scanning the photoelectron intensity along the $k_y$ direction while maintaining a fixed $k_x$, effectively cutting directly through the $K$ ($K'$) high-symmetry points of the hexagonal Brillouin zone (see Sec. 3 of the Supplementary Material for details~\cite{SM}). Notably, the observed band profile cleanly resolves the linear dispersion, which surprisingly exhibits a clear formulation of a ``camel-back" structure associated with a finite gap of $\sim 0.2$ eV opening at the Dirac point---strikingly reminiscent of bilayer graphene. The Fermi energy ($E_F$) is found to locate approximately $0.1\,\text{eV}$ above the bottom of the conduction band, corroborating the electron-doping (see table~\ref{tab:carrier_properties}) nature of the system.

The distorted $E-k$ dispersion, characterized by the distinct camel-back structure in the conduction and valence band, is modeled using Eq.~(\ref{eq:deformedDirac}). To accurately capture this strong distortion resulting from the interaction with the underlying buffer layer, we employ the exact low-energy band formulation derived from the conventional bilayer graphene model~\cite{CastroNeto2009}:
\begin{equation*}
E_{\pm}(k)
= \pm \sqrt{\Delta_0^2 + \frac{\gamma_1^2}{2} + (v_F \hbar k)^2 -
\Lambda_k},
\end{equation*}
with the inner term $\Lambda_k$ defined as
\begin{equation}
\label{eq:deformedDirac}
\Lambda_k
= \sqrt{\frac{\gamma_1^4}{4} + (v_F \hbar k)^2 (\gamma_1^2 +
4\Delta_0^2)}.
\end{equation}
where the upper and lower signs ($\pm$) denote the conduction and valence bands, respectively. Here, $v_F$ is the bare Fermi velocity, $\hbar k$ is the in-plane momentum measured relative to the $K$ ($K'$) point, and $\Delta_0$ represents the sublattice asymmetry gap.
The parameter $\gamma_1$ denotes the effective interlayer coupling energy, which rigorously reproduces the camel-back curvature by accounting for the strong electronic interaction between the epitaxial graphene monolayer and the underlying buffer layer.

This exact analytical expression perfectly captures the strongly distorted Dirac cone and the microscopic ``camel-back'' structure observed in our ARPES measurements. A detailed discussion of this formulation, including its exact derivation from the full four-band Hamiltonian and the extraction of the fitting parameters, is provided in Sec.~S3 of the Supplementary Material~\cite{SM}.
%

%

A red solid line in Fig.~\ref{fig:ARPES} is thus obtained as the best fit to the observed distorted Dirac dispersion, yielding the final parameters of $v_F = 1.06 \times 10^6\,\text{m/s}$, $\Delta_0 = 0.1\,\text{eV}$, and $\gamma_1 = 0.39\,\text{eV}$. 
Here, the universal interlayer coupling value of standard bilayer graphene ($\gamma_1 = 0.39\,\text{eV}$) is utilized as an effective parameter to mathematically reproduce the severe intrinsic band-warping~\cite{CastroNeto2009}. 
Crucially, within this effective bilayer-derived formulation, the bare sublattice potential asymmetry parameter $\Delta_0$ directly governs the energy gap opening at the Dirac point, defined as $E_g = 2\Delta_0$~\cite{Mucha2010}. The extracted value of $\Delta_0 = 0.1\,\text{eV}$ translates to a bare zero-field band gap of $0.2\,\text{eV}$. This exhibits excellent quantitative consistency with the clear separation between the conduction and valence bands directly resolved in the ARPES intensity map (Fig.~\ref{fig:ARPES}).

It is noteworthy that the extracted gap size of $0.2\,\text{eV}$ is exceptionally large compared to typical values reported for conventionally gated exfoliated graphene. 
In epitaxial graphene grown on a SiC substrate by thermal decomposition, it is well established that the strong interfacial interaction with the underlying buffer layer inherently breaks the A-B sublattice symmetry, providing an initial energy gap at the Dirac point. Notably, a previous study reported a similar gap size of $\sim$0.26 eV~\cite{Zhou2007}.
Theoretically, as demonstrated by Luk'yanchuk and Bratkovsky, the discrete honeycomb lattice inherently breaks continuous relativistic invariance, providing a fundamental orbital gap driven by sublattice potential asymmetry ($\Delta$)---precisely what we directly capture via our ARPES measurements \cite{Lukyanchuk2008}. 

\subsection{Landau Level Fan Chart and Optical Transitions}

The comprehensive set of microscopic band parameters obtained from our ARPES analysis provides a robust foundation for predicting the magnetic-field evolution of this system. In our zero-field ARPES analysis, the severe band distortion and the macroscopic breakdown of electron-hole symmetry---driven by the built-in electron doping---were effectively captured by a modified full-band model incorporating an effective coupling parameter ($\gamma_1$). However, directly applying this bilayer-derived formulation to calculate the Landau levels (LLs) of our strictly monolayer system is physically inappropriate. 

Instead, we must consider how this intrinsic band-warping manifests in the presence of a perpendicular magnetic field. The strongly distorted Dirac dispersion implies the emergence of a giant orbital magnetic moment. In our effective LL formulation, this profound consequence of broken electron-hole symmetry is mapped onto a strong linear magnetic field coupling term ($\kappa B$) acting specifically on the zero-mode states. 
To precisely evaluate the evolution of the LLs, we numerically diagonalize the full effective Hamiltonian incorporating this $\kappa B$ term. The detailed matrix formulation of the effective Hamiltonian and its exact diagonalization are provided in the Appendix.

\begin{figure}[htbp]
\centering
\includegraphics[width=0.7\columnwidth]{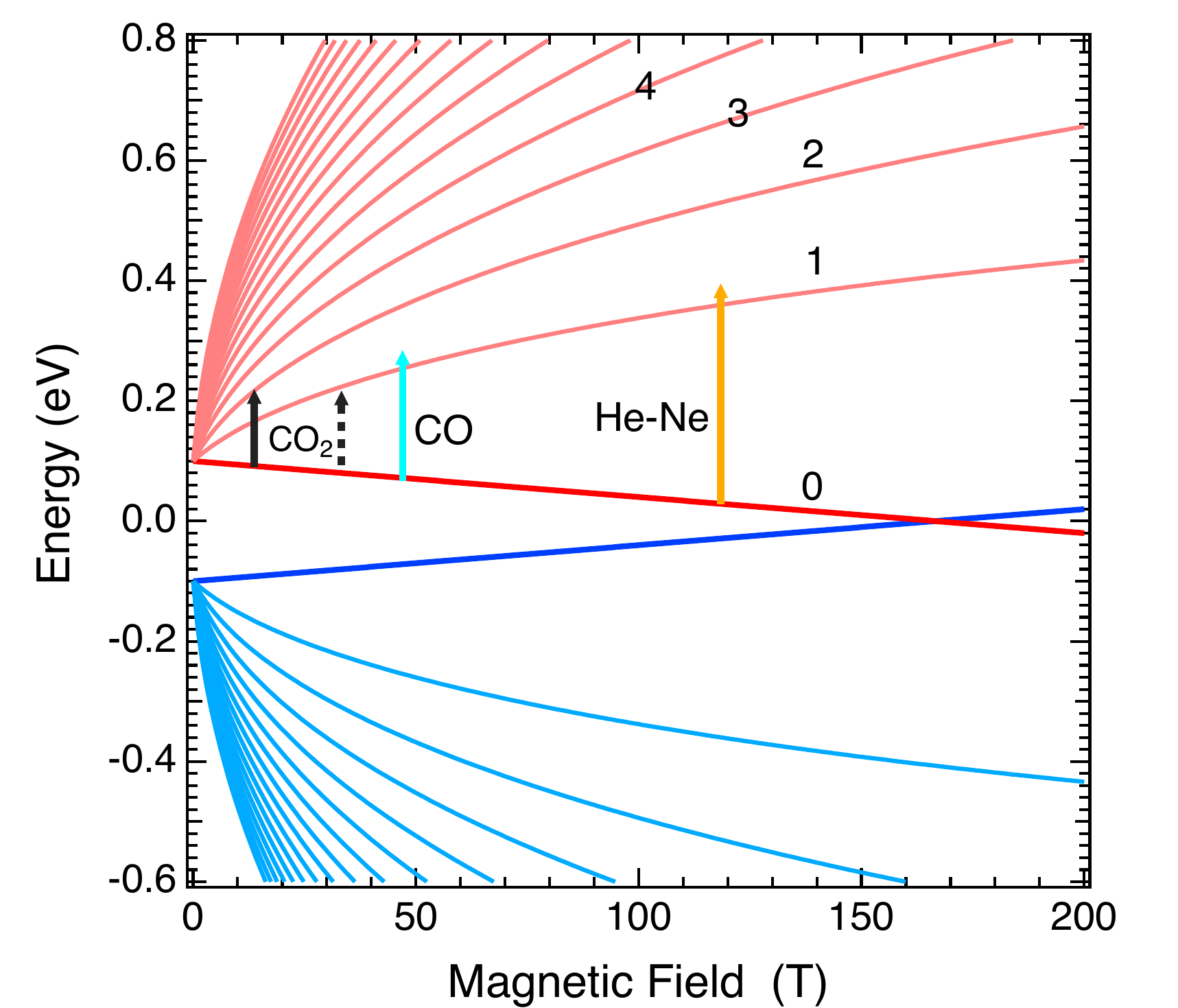}
\caption{\label{fig:LLFabMain}
\footnotesize
Calculated Landau-level (LL) fan chart up to 200~T based on the distorted Dirac dispersion model, using parameters extracted from our ARPES analysis ($v_F = 1.06 \times 10^6\,\text{m/s}$, 
$\Delta = 0.1\,\text{eV}$). The red and blue curves represent the LLs for the conduction and valence bands, respectively. The solid vertical arrows (black, blue, and orange) indicate the optical transition energies corresponding to the STC experiments shown in Fig.~\ref{fig:waveformSTC}(a) for the CO$_2$ ($\hbar\omega_{0} = 0.130\,\text{eV}$), CO ($\hbar\omega_{0} = 0.218
\,\text{eV}$), and He-Ne ($\hbar\omega_{0} = 0.367\,\text{eV}$) lasers. Notably, the dashed arrow indicates where the CO$_2$ laser transition would be expected to appear if it were a conventional cyclotron resonance (CR). A level crossover between the $0^+$ (conduction band) and $0^-$ (valence band) LLs occurs at approximately 160~T.
}
\end{figure}

A profound and highly intriguing feature emerges when directly correlating this calculated single-particle LL topology with our experimental magneto-optical spectra. In the lower-field regime below 100~T (at 15~T and 47~T), the transmission dips retain a conventional CR-like profile. However, as shown in Fig.~\ref{fig:LLFabMain}, the calculated transition energies exhibit a definite disagreement with the observed low-field CR positions, assigning them not to the expected $0^+ \to 1^+$ transition, but rather to the $0^+ \to 2^+$ transition. 
Furthermore, the transitions observed with the CO and He-Ne lasers also exhibit a subtle discrepancy with the expected $0^+ \to 1^+$ CR transition energies. Taken together, these observations strongly suggest that all the absorption spectra presented in Fig.~1 do not originate from conventional CR, but are instead driven by collective plasma excitations such as helicon waves.

Crucially, this rigorous theoretical treatment reveals a striking topological consequence: driven by the substantial $\kappa B$ term, the $N=0^+$ and $N=0^-$ states rapidly converge. 
The calculation predicts that these zero-mode levels will inevitably cross at a certain extreme magnetic field, naturally setting the stage for exploring entirely new physical phenomena in the ultrahigh-field regime.
A dramatic optodynamic transformation is expected to occur upon surpassing the critical field threshold of $\sim$160~T. 
Exactly where the fan chart dictates the unconventional level-crossing and subsequent energy inversion between the $0^+$ (CB) and $0^-$ (VB) levels---driving the system from a positive to a negative gap state at higher fields---the CR-like features might entirely vanish from the experimental spectra, transitioning into an exceptionally broad, featureless absorption profile. 

This spectral collapse suggests a highly fascinating possibility: a magnetic-field-induced phase transition where the level inversion forcibly mixes the electron and hole degrees of freedom, driving the system into a compensated Dirac metallic state. 
Following this crossover, the populations of electrons and holes become nearly balanced. Consequently, the dominant optical response shifts from a collective excitation purely of electrons (observed as the CR-like dips at lower fields) to a fully cooperative collective excitation involving both electrons and holes simultaneously. This perfectly compensated electron-hole plasma provides the ideal environment for the emergence of macroscopic, magnetohydrodynamic collective modes, namely Alfv\'en waves. To rigorously verify this compelling scenario, it is indispensable to examine the ultimate high-field data in the 500-T regime by employing a collective Alfv\'en wave propagation analysis, which we discuss in detail in the following section.

\begin{figure}[htbp]
\centering
\includegraphics[width=0.8\columnwidth]{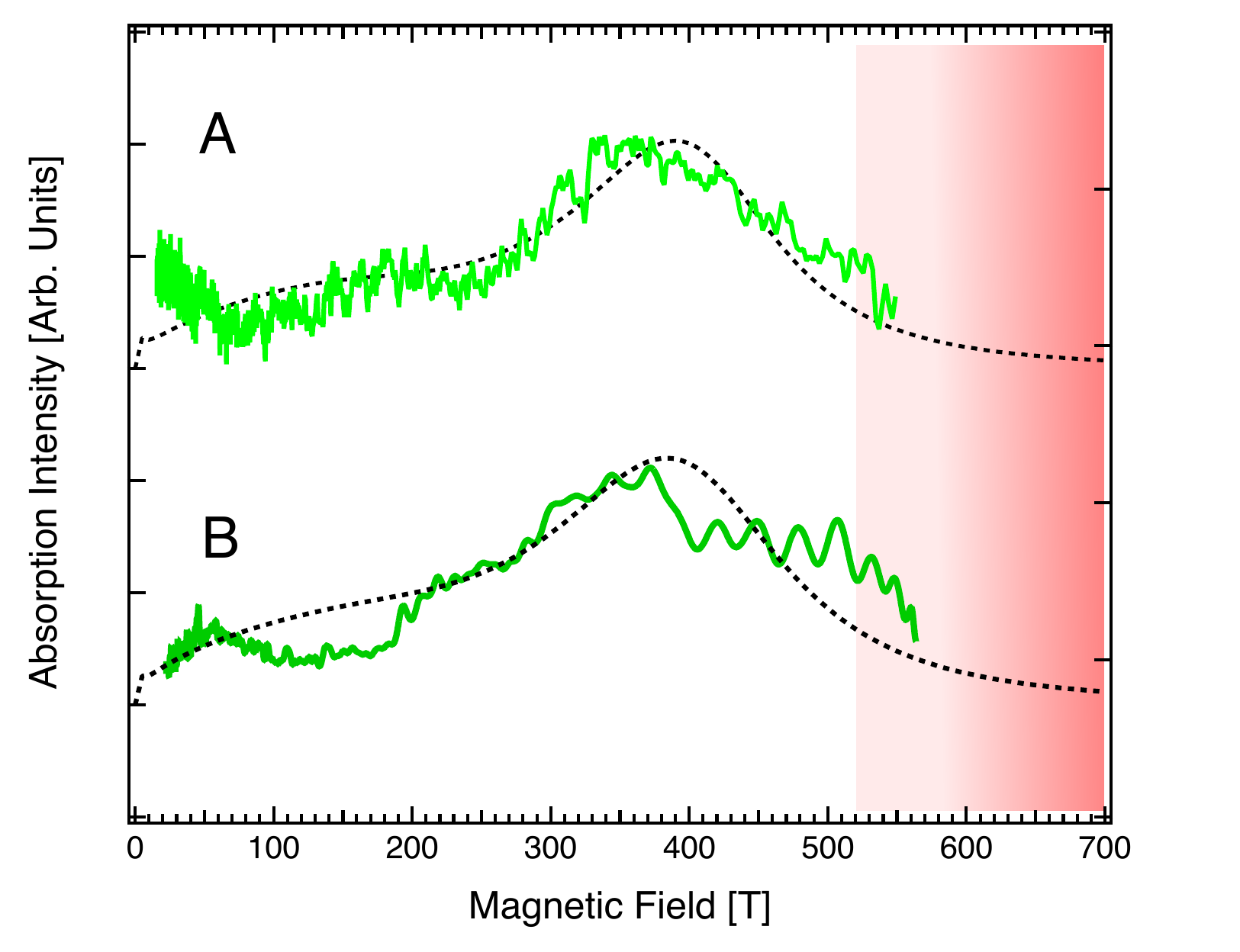}
\caption{\label{fig:Alven600T}
\footnotesize Magneto-infrared absorption spectra for A: Sample A and B: Sample B in ultrahigh magnetic fields up to 560~T, measured at room temperature using a thulium fiber laser (photon energy: 0.636~eV). The solid curves represent the optical absorption spectra calculated via the collective Alfv\'en wave model. The red-shaded graduation above 500~T marks the region that is interpreted as being potentially influenced by stray infrared radiation emitted from the approaching liner of the EMFC (Detailed experimental configurations are provided in the Supplementary Material---Sec. S1~\cite{SM}).}
\end{figure}


\subsection{IR Magneto-photo Absorption in EMFC Measurements  and Alfv\'en Wave Analyses}

Figure~\ref{fig:Alven600T} displays the high-energy magneto-photo absorption spectra capturing the cyclotron resonance (CR) regimes under ultrahigh magnetic fields ranging from 300 to 600~T. The measurements were performed using a thulium fiber laser operating at a wavelength of 1.950~$\mu$m, which corresponds to a photon energy of 0.636~eV. For both Sample A and Sample B, broad absorption peaks accompanied by several oscillatory structures are clearly observed over a wide magnetic field range from 100 to 560~T. 
It should be noted that these raw spectral profiles are identical to those presented in Fig.~3 of Ref.~\cite{Nakamura2020}. However, as previously addressed in the discussion regarding Fig.~\ref{fig:waveformSTC}, the vertical axis in the previous report was excessively expanded, masking the macroscopic baseline behavior. 
Note that at approximately 160-200~T, a shoulder like peak appears prior to the main peak around 400 T, corresponding to a prominent topological crossing between the $0^+$ (conduction band) and $0^-$ (valence band) LLs (see Fig.~\ref{fig:LLFabMain}).

 The black dashed curves in Fig.~\ref{fig:Alven600T} exhibits the theoretical optical absorption calculated via the collective Alfv\'en wave model~\cite{bassani1975solids} [Equation~(S5) in Supplementary Material---Sec.~S4~\cite{SM}]. 
 The experimental waveforms are reproduced by the physical parameters extracted from the fitting, and the results are listed in Table~\ref{tab:alfven_parameter}.

\begin{table}[htbp]
\centering
\caption{\label{tab:alfven_parameter}
Fitting parameters for the Alfv\'en wave analysis. The total spectral broadening $\Gamma_B$ and the quantum scattering lifetime $\tau$ are also shown.
}
\begin{tabular}{lccccc}
\hline\hline
Sample & $v_F$ ($10^6$~m/s) & $\Delta_B$ (eV) & $E_F$ (meV) & $\Gamma_B$ (eV) & $\tau$ (fs) \\
\hline
\textbf{Sample A} & 0.97 & 0.12 & 40  & 0.14 & 4 \\
\textbf{Sample B} & 0.86 & 0.12 & 100 & 0.13 & 4 \\
\hline\hline
\end{tabular}
\end{table}
 

These best-fit parameters exhibit an outstanding agreement with the intrinsic band parameters independently determined by the zero-field ARPES measurements in Sec.~\ref{sub:arpes} ($v_F = 1.06 \times 10^6$~m/s, $\Delta_0 =0.1$~eV). 
The Fermi energy $E_F$ for each sample is almost identical to the value determined from the transport measurements (listed in Table~\ref{tab:carrier_properties}). 
This quantitative consistency provides unequivocal evidence that the characteristic magneto-absorption features observed in Fig.~\ref{fig:Alven600T} represent absolutely nothing other than the collective Alfv'en wave propagation.

 The Landau-level broadening lifetime $\tau$ in Eq.~(S12) was optimized for fitting; the obtained scattering lifetime of $\tau \approx 4$~fs is remarkably short compared to the values typically reported for graphene (10$\sim$20~fs)~\cite{Jiang2007}. The mobilities listed in Table~\ref{tab:carrier_properties} lead to the value $\tau \approx 12$~fs for both samples.
In sharp contrast to conventional semiconductors---where magnetic fields suppress scattering via spatial localization---the scattering rate in this system accelerates dramatically above 160~T due to the crossover into a perfectly compensated electron-hole plasma regime. The equal coexistence of dense electrons and holes opens an intense electron-hole Coulomb scattering channel, which is further amplified by the enhanced density of states of the compressed, camel-back Dirac topology. Notably, the robust survival of macroscopic Alfv\'en oscillations up to 600~T despite this ultra-fast scattering provides definitive proof of the coherent, collective nature of the Alfv\'en mode, which inherently outlasts individual particle lifetimes. Furthermore, this macroscopic gap expansion under megagauss fields provides compelling evidence for a many-body phase transition driven by "magnetic catalysis." Under extreme spatial confinement within a sub-nanometer magnetic length ($l_{B} = \sqrt{\hbar/eB} \approx 1.8$--$1.0$~nm for $B = 200$--$600$~T), the unscreened Coulomb interaction is strongly enhanced. 

Historically, the behavior of the zero-mode Landau levels (LLs) under magnetic fields has been heavily investigated in the context of magnetic catalysis. As predicted by Gusynin \textit{et al.}, the distinct chiral properties of Dirac fermions make the zero-energy state highly vulnerable to many-body correlation effects, which are expected to dynamically generate or widen energy gaps \cite{Gusynin2005}. Extensive experimental efforts have sought to observe this field-induced gap generation in systems lacking a substantial zero-field gap. For instance, Giesbers \textit{et al.} observed only a small gap in exfoliated graphene on SiO$_2$ due to disorder suppression \cite{Giesbers2009}, while Young \textit{et al.} \cite{Young2012} and Henriksen \textit{et al.} \cite{Henriksen2010} observed anomalously large, yet still perturbative, interaction-driven energy shifts (e.g., yielding gap openings of $\sim 20\,\text{meV}$ up to $31\,\text{T}$). 
In all these conventional cases, the magnetic field acts to split the degenerate $N=0$ levels, thereby opening or widening an energy gap. 
Furthermore, these experimental signatures of field-driven gap enhancements are fundamentally supported by extensive theoretical frameworks predicting interaction-driven mass generation under magnetic fields \cite{Fuchs2007, Yang2007}.

In stark contrast, our system operates under an entirely different topological paradigm. We begin with a giant pre-existing zero-field gap ($2\Delta_0 \sim 0.2$ eV). When subjected to extreme magnetic fields, rather than undergoing further splitting or gap widening, the strong linear magnetic field coupling ($\kappa B$)---originating from the giant orbital magnetic moment of the distorted Dirac cone---drives the $N=0^+$ and $N=0^-$ levels to rapidly converge. This topological trajectory dictates that the initial energy gap is not amplified, but rather closes and ultimately inverts. This critical level crossover transitions the system into an fully compensated electron-hole plasma regime at ultrahigh magnetic fields, setting an unprecedented stage for our megagauss magneto-optical investigations. Furthermore, previous studies indicate that robust many-body interactions require high sample quality \cite{Giesbers2009}. Our epitaxial graphene samples exhibit sufficiently high carrier mobility (see Table~1) to preserve these strong interactions. Once the $N=0$ levels invert and the system enters the collective Alfv\'en wave regime (200--560 T), our analysis reveals a dynamically enhanced parameter $\Delta_B > \Delta_0$. Crucially, in this post-crossover plasma phase, $\Delta_B$ no longer represents a simple energy gap. Instead, this remarkable increase in $\Delta_B$ indicates that the strong interactions between the coexisting electrons and holes strictly amplify the fundamental sublattice potential asymmetry of the system. Thus, our results highlight a novel macroscopic manifestation of many-body effects: not the opening of a simple gap, but the dynamic enhancement of sublattice asymmetry driven by a magnetic-field-induced electron-hole plasma.

\begin{figure}[htbp]
\centering
\includegraphics[width=0.9\columnwidth]{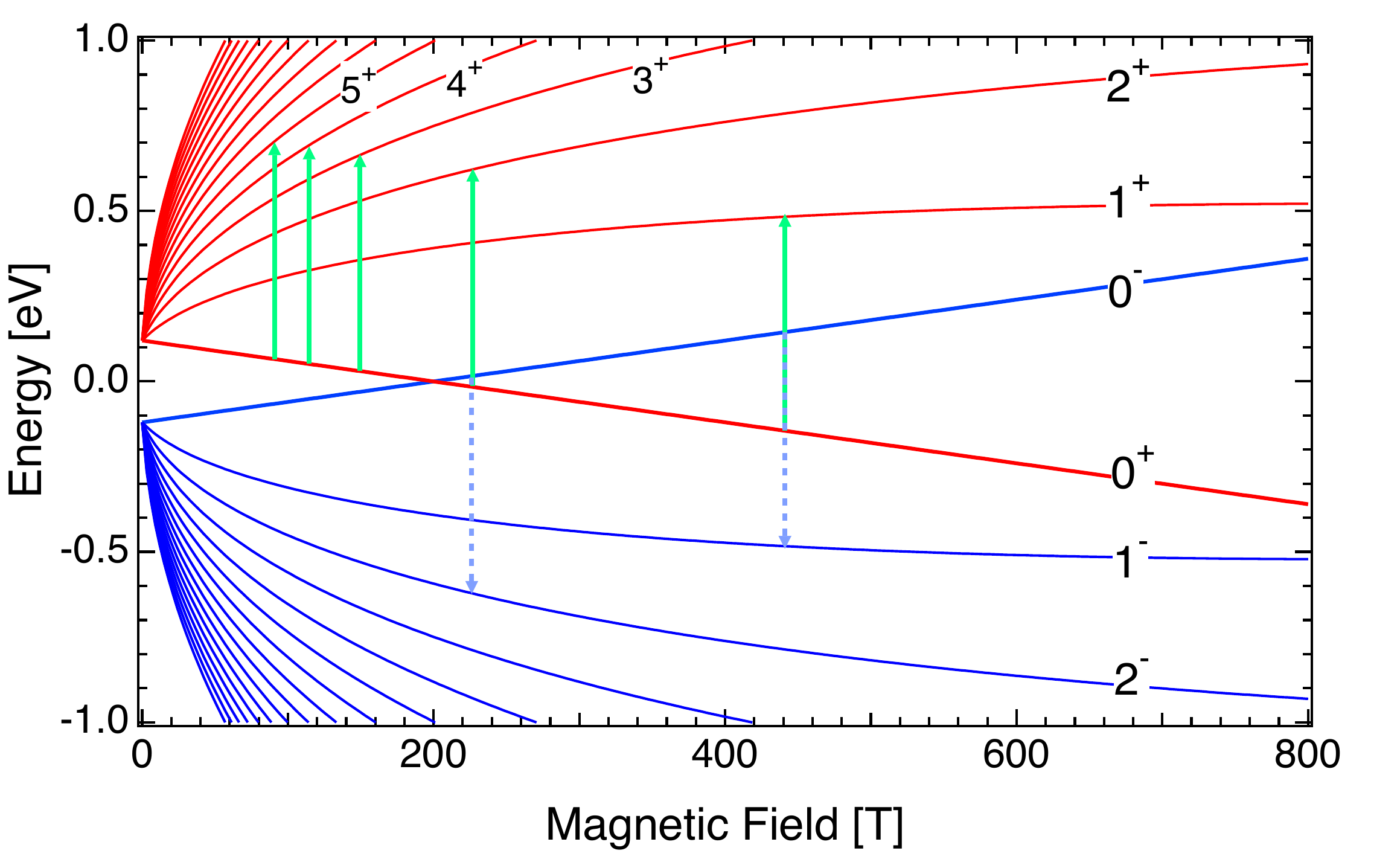}
\caption{\label{fig:LLFanCh800T}
Landau-level (LL) fan chart calculated up to 800~T. The upper and lower manifolds correspond to electron and hole LLs, respectively. Vertical arrows (solid and dashed) denote resonant optical transitions matching the thulium fiber laser photon energy (0.636~eV). The $0^+$ and $0^-$ levels undergo a crossover at $B = 200$~T. Note that this fan chart slightly differs from Fig.~\ref{fig:LLFabMain}, as it employs the parameters derived from the high-field Alfv\'en wave analysis.
}
\end{figure}

Figure~\ref{fig:LLFanCh800T} presents the LL fan chart calculated up to 800~T. The calculation is based on the band parameters optimized for the magneto-optical spectra in ultrahigh magnetic fields; specifically, we utilized the parameter set listed in Table~\ref{tab:alfven_parameter} obtained during the fitting process of the Alfv\'en wave spectral curves of Sample A in the 150--560~T range. The crossover between the $0^+$ and $0^-$ levels takes place at $B \sim 200$~T. 

It is crucial to physically distinguish the parameter $\Delta_B$ utilized in this high-field calculation from the initial parameter $\Delta_0$ obtained via the ARPES measurements. The parameter $\Delta_0$ represents the ``bare'' sublattice potential asymmetry intrinsic to the system at zero magnetic field. In contrast, $\Delta_B$ serves as a ``renormalized'' parameter that inherently incorporates the strong many-body correlation effects emerging within the ultrahigh-field electron-hole plasma phase. Therefore, the observed enhancement $\Delta_B > \Delta_0$ is not a simple widening of a conventional energy gap. Rather, this quantitative difference directly measures the robust electron-hole interaction strength that dynamically amplifies the fundamental sublattice asymmetry in the post-crossover regime. Consequently, the apparent zero-field energy gap ($E_g = 2\Delta_B$) depicted in this fan chart, Fig.~\ref{fig:LLFanCh800T} is entirely fictitious. It represents a backward extrapolation from the ultrahigh-field state, serving purely as an effective mathematical baseline necessary to accurately compute the heavily dressed LLs in the megagauss plasma regime.

The solid green and dashed blue arrows indicate resonant optical transitions that quantitatively match the 0.636~eV photon energy of the thulium fiber laser. As clearly illustrated, both electron and hole transitions become simultaneously allowed only after the crossover of the $0^+$ and $0^-$ levels. Prior to this crossover, the allowed optical excitations are strictly limited to electron transitions into higher LLs.
The transitions at 440~T and 220~T directly correspond to the main absorption peak and its shoulder structure, respectively, as visualized in Fig.~\ref{fig:Alven600T}. The weak absorption at 220~T occurs because the system has just undergone the level crossover; here, the electron population still overwhelmingly dominates over the hole population, preventing the full establishment of the collective Alfv\'en mode.
%


This physical picture marks a profound departure from conventional metal-like plasmas governed by Kohn's theorem~\cite{Hartnoll2007}. While standard Galilean-invariant metals host a single-component plasma comprised solely of electrons---where the strict proportionality between charge current and momentum prevents the decay of current without external scatterers---the magnetic-field-induced crossing of the conduction and valence bands in our monolayer graphene establishes a two-component, perfectly compensated electron-hole plasma.
This charge-neutral, symmetric Dirac fluid decouples the electrical and momentum currents, allowing the system to behave as a non-degenerate relativistic gas. Such a symmetric fluid serves as a laboratory-scale analog to the relativistic electron-positron plasmas found in astrophysical extremes---such as pulsar magnetospheres and the early universe---as well as the quark-gluon plasmas (QGP) governing high-energy nuclear physics~\cite{Hartnoll2007, Muller2009}. 

In fact, the emergence of this collective mode is even more striking considering that, in our system, the exceptionally short scattering lifetime ($\tau \approx 4$~fs) initially places this Dirac fluid in a highly dissipative, strongly coupled regime where the hydrodynamic cyclotron mode is severely overdamped by impurity scattering~\cite{Hartnoll2007}.
Under typical low-to-moderate magnetic fields, this extreme dissipation forces the system to manifest only as a featureless, highly damped Drude-like plasma absorption background (as reminiscent of the broad spectral profiles in Fig.~1(b)). 
The true milestone of our study is the demonstration that under unprecedented, megagauss-class magnetic fields, the macroscopic magnetic tension of this otherwise dissipative 2D plasma is dynamically stiffened to such an extreme degree that the collective resonance frequency ($\omega_c$) successfully overcomes the massive scattering rate ($1/\tau \ll \omega_c$). This enables the long-sought collective hydromagnetic mode to finally emerge as spectacular, well-defined resonance peaks systematically spanning the ultrahigh-field range of 200---600~T. 

Before concluding, it is crucial to address the physical origin of this exceptionally robust band deformation. While the substrate-induced interaction essentially breaks the sublattice symmetry, we propose that the rigid maintenance of this heavily distorted Dirac dispersion---even under megagauss fields---is fundamentally stabilized by many-body band renormalization. 
As indicated by the zero-field transport measurements (Table~\ref{tab:carrier_properties}), the initial built-in electron doping ($n \sim 10^{11}\,\text{cm}^{-2}$) completely breaks the electron-hole symmetry. This inherently places the system in a weakly screened, highly interacting regime where strong Coulomb interactions rigidly lock the high-momentum linear dispersion and the camel-back structure. It is precisely this robust, interaction-driven ground state that allows the system to host the stable collective Alfv\'en modes up to the 600~T regime. 

The observed ultrahigh-field magneto-optical profiles thus provide a unique, laboratory-scale window into the fundamental dynamics of symmetric relativistic fluids, beautifully unifying tabletop solid-state spectroscopy with the physics of high-energy and astrophysical extremes~\cite{Hartnoll2007, Muller2009}. 
On a more profound level, this rigidly distorted Dirac dispersion evokes a striking analogy to the geometry of relativistic spacetime. While the ideal graphene band structure forms a perfectly symmetric Minkowski-like cone---analogous to the pristine light cone in flat spacetime ($x^2+y^2+z^2-c^2t^2=0$)---our heavily interacting, symmetry-broken system strongly distorts this pristine geometry. Just as the presence of mass and energy warps flat Minkowski space into a curved spacetime in Einstein's general relativity, the strong many-body interactions and broken symmetries in our Dirac fluid effectively ``warp" the ideal massless cone into a robust, gapped hyperboloid-like structure. Observing such a fundamental geometric deformation persisting up to the megagauss regime provides a fascinating, tangible solid-state analogue to the spacetime distortions found in extreme astrophysical environments.

\section{Summary and Conclusion}
In summary, based on ARPES measurements of $n$-doped graphene ($n=3\text{--}7 \times 10^{11} \text{ cm}^{-2}$), we revealed that the Dirac electron-hole symmetry is fundamentally broken in this system, resulting in a camel-back distorted Dirac dispersion. We demonstrated that a level crossing between the electron $n=0^+$ and hole $n=0^-$ LLs occurs at 200~T, driving the system from a positive to a negative gap state at higher fields. Furthermore, by applying the band dispersion parameters derived from ARPES, we analyzed the magneto-infrared photo-absorption spectra up to ultrahigh magnetic fields of 560~T. 
We discovered that the optical response in this extreme field regime is not governed by conventional single-electron cyclotron resonance (CR), but rather by collective Alfv\'en waves originating from the electron-hole plasma. 

Through the combined analysis of the zero-field ARPES and high-field Alfv\'en wave data, we comprehensively determined the microscopic band dispersion parameters across entirely different physical regimes. We found that while the Fermi velocity $v_F$ remains remarkably robust under extreme magnetic fields---preserving the fundamental high-momentum linear dispersion---the sublattice potential asymmetry parameter $\Delta$ is distinctly enhanced. Furthermore, the intrinsic band distortion characterized by $\gamma_1$ at zero field is effectively encapsulated by the strong linear magnetic field coupling term ($\kappa B$) introduced in our ultrahigh-field effective Hamiltonian. Driven precisely by this $\kappa B$ term, the giant initial energy gap does not widen; rather, the $0^+$ and $0^-$ levels rapidly converge, close, and ultimately invert. Consequently, the pronounced increase in $\Delta$ ($\Delta_B > \Delta_0$) extracted in the ultrahigh-field regime reflects a macroscopic many-body effect: a dynamic amplification of the fundamental sublattice asymmetry, driven strictly by the strong Coulomb interactions within the newly established two-component electron-hole plasma.

\appendix
\setcounter{equation}{0}
\renewcommand{\theequation}{A\arabic{equation}}
\setcounter{figure}{0}
\renewcommand{\thefigure}{A\arabic{figure}}
\section*{\normalsize \bfseries Appendix: Landau Level Fan Chart Calculations}

Utilizing the robust band parameters extracted from our ARPES analysis ($v_F$ and $\Delta_0$), we construct the relativistic Landau level (LL) fan chart under ultrahigh magnetic fields. 

In the megagauss regime, simple analytical approximations are insufficient because they fail to fully account for the pronounced valley splitting, anomalous energy shifts, and severe level mixing (anti-crossings) among different quantum numbers. To accurately capture these complex behaviors---particularly the anomalous level-crossing driven by the giant orbital magnetic moment associated with the gapped Dirac topology---we explicitly introduce a linear magnetic field term $\kappa B$ into the effective model. Therefore, to construct the precise LL fan chart presented in this study, we numerically diagonalize the full matrix representation of the Hamiltonian $H_{\lambda}(B)$:

\begin{equation}
H_{\lambda}(B) = 
\begin{pmatrix}
E_{0, \lambda}^0 & \kappa B & 0 & 0 & \cdots \\
\kappa B & E_{1, \lambda}^0 & \beta\sqrt{2\cdot B} & 0 & \cdots \\
0 & \beta\sqrt{2\cdot B} & E_{2, \lambda}^0 & \beta\sqrt{3\cdot B} & \cdots \\
0 & 0 & \beta\sqrt{3\cdot B} & E_{3, \lambda}^0 & \cdots \\
\vdots & \vdots & \vdots & \vdots & \ddots
\end{pmatrix},
\end{equation}
where the band index $\lambda \in \{\text{CB}, \text{VB}\}$ explicitly distinguishes the conduction band ($\lambda = \text{CB}$) and the valence band ($\lambda = \text{VB}$) manifolds. The matrix elements are directly linked to our ARPES-derived fundamental values ($v_F$, $\Delta_0$) and the linear coupling parameter $\kappa$ through the following physical relations:

\begin{itemize}
    \item $E_{n, \lambda}^0$: The index-dependent diagonal base energy for the $n$-th Landau level, incorporating the site-energy asymmetry ($\Delta_0$) and the linear magnetic field shift. Their explicit diagonal forms are expressed as:
    \begin{equation}
    E_{n, \text{CB}}^0 = \Delta_0 + \kappa \cdot n \cdot B
    \end{equation}
    \begin{equation}
    E_{n, \text{VB}}^0 = -\Delta_0 - \kappa \cdot n \cdot B
    \end{equation}
    
    \item $\kappa$: The coefficient dictating the direct linear magnetic field coupling ($\kappa B$). This parameter phenomenologically captures the anomalous energy evolution (such as the zero-mode level crossing), which physically originates from the Berry curvature and the resulting giant orbital magnetic moment at the band edges of the symmetry-broken Dirac cone.
    
    \item $\beta$: The effective Dirac coupling constant regulating the relativistic $\sqrt{B}$ scaling of the inter-LL mixing, defined precisely by the ARPES Fermi velocity $v_F$ and the elementary charge $e$:
    \begin{equation}
    \beta = v_F \sqrt{2e\hbar}
    \end{equation}
\end{itemize}

\noindent
The diagonalization of this comprehensive matrix yields the exact eigenvalues that intricately capture the severe level mixings and anomalous energy shifts, thereby allowing us to accurately trace the resonant optical transition pathways observed in our extreme high-field magneto-optical spectra without relying on restrictive analytical approximations.

Crucially, it is the presence of this linear term $\kappa B$ that overcomes the initial energy gap $2\Delta_0$ and dictates the anomalous energy evolution of the zero-mode Landau levels. As a direct consequence, this term induces the critical level crossover between the $0^+$ and $0^-$ states observed in the ultrahigh-field regime. Physically, the linear term $\kappa B$ driving the $N=0$ crossing fundamentally originates from the band-warping parameter (interlayer interaction) of the highly distorted Dirac cone.

Furthermore, this physical picture is in excellent agreement with the rigorous theoretical framework established by Ando and Suzuura~\cite{AndoSuzuura2017}. In their work, they demonstrated that higher-order band-warping effects and structural asymmetries in modified Dirac systems inherently induce anomalous energy shifts and splittings of the zero-mode Landau levels. In this context, our effective linear term $\kappa B$ serves as a robust macroscopic manifestation of these microscopic interband interactions, thereby solidly validating the theoretical mechanism behind the field-induced crossover between the $0^+$ and $0^-$ states.

\vspace{20pt}



\vspace{20pt} 


\clearpage

\begin{center}

{\large \bfseries Supplemental Material: \\ 
From Zero to Mega-Gauss Fields: Comprehensive Magnetophotonic Spectroscopy of Graphene Dirac Cones}
\end{center}
\vspace{2ex}

\setcounter{equation}{0}
\setcounter{figure}{0}
\renewcommand{\theequation}{S\arabic{equation}}
\renewcommand{\thefigure}{S\arabic{figure}}


\noindent{\normalsize \bfseries S1. Sample Setting for IR and FIR Optical Measurements}
\vspace{2ex}

\begin{figure}[htbp]

\includegraphics[width=0.8\columnwidth, angle=0]{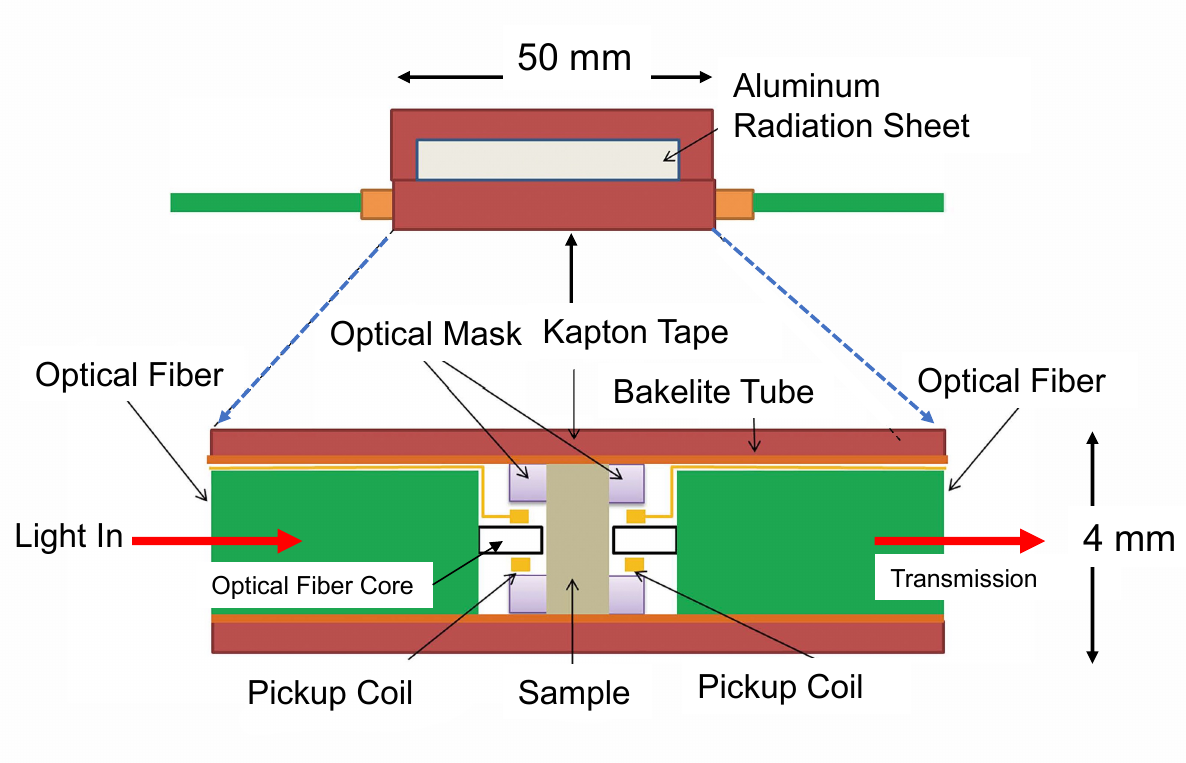}
\centering
 \caption{\label{fig:SampleSet}
Design of the sample holder used for infrared optical transmission measurements during the EMFC experiments. To ensure highly efficient optical transfer, a large-core optical fiber with a diameter of 1.0--1.5~mm and a total length of 10--30~m is utilized.}
\end{figure}
Special care must be taken during sample installation for infrared (IR) optical transmission measurements under the electromagnetic flux compression (EMFC) experiment configuration~(see details Ref.~\cite{Takeyama2026}. First, as the imploding high-temperature liner (which compresses the magnetic flux) approaches the sample holder, it generates intense broadband FIR and IR thermal irradiation. To shield the detector from this intense stray light, a dedicated multi-layered radiation shield was constructed around the 4-mm-diameter sample holder as illustrated in Figure~\ref{fig:SampleSet}. A 5-cm-wide Kapton tape was prepared with a length sufficient for 2--3 wraps around the holder. Onto this Kapton base, a super-radiation shield sheet (with an aluminum layer thinner than 300~nm) was affixed, leaving a lateral margin of approximately 5~mm and a circumferential margin of 1~cm at each end to prevent electrical shorting during the implosion. This composite sheet was then tightly wrapped around the sample holder 2--3 times. Furthermore, because FIR and IR photons are highly prone to diffraction and leakage around the sample edges, meticulous optical masking was applied to ensure the integrity of the transmission path.

\vspace{2ex}
\noindent{\normalsize \bfseries S2. Cyclotron Resonance Optical Absorption Spectra of Graphene}
\vspace{2ex}

To simulate the line shape of the bare, non-interacting single-particle cyclotron resonance (CR), we utilized the analytical expression for the optical absorption $A(B)$ as a function of the magnetic field $B$. This formulation is rigorously derived from the Kubo formalism for 2D Dirac fermions, following the framework presented in Appendix (A37) of Booshehri \textit{et al.} \cite{Booshehri2012}. By adapting their frequency-dependent conductivity expression to our fixed-photon-energy and field-sweeping configuration, the field-dependent absorption is explicitly modeled as:
\begin{equation}
A(B) = \alpha_{\mathrm{eff}}\frac{\frac{\sqrt{B}}{\tau}}{\left(\sqrt{\frac{B_c}{B}} - 1 \right)^2 + \left( \frac{1}{\tau}\right)^2}
\end{equation}
where $A(B)$ represents the optical absorption as a function of the magnetic field $B$, $B_c$ denotes the resonance magnetic field, and $\tau$ is the dimensionless phenomenological scattering lifetime of the Dirac fermions. 
To ensure that the peak absorption at the resonance condition ($B = B_c$) remains strictly locked to the universal interband absorption baseline of monolayer graphene ($\pi\alpha_{\mathrm{universal}} \approx 2.29\%$) regardless of the damping strength $\tau$, we define the effective coupling coefficient as:
\begin{equation}
\alpha_{\mathrm{eff}} = \frac{\pi\alpha_{\mathrm{universal}}}{\tau}
\end{equation}
where $\alpha_{\mathrm{universal}} \approx 1/137$ is the fine-structure constant. This renormalization ensures that the spectral line shape naturally broadens as the scattering increases (smaller $\tau$) while keeping the maximum absorption depth physically restricted to the universal single-layer limit.

The magneto-transmission spectrum $T(B)$ is subsequently obtained via the linear relation:
\begin{equation}
T(B) \approx 1 - A(B)
\end{equation}
This simplified linear boundary condition is strictly justified for an atomically thin two-dimensional crystal like monolayer graphene. Unlike bulk (3D) materials where optical transmission follows the exponential Beer-Lambert law ($T = e^{-\alpha d}$), a 2D atomic sheet acts as a discrete interface where incoming photons undergo a single-pass interaction. Given that the optical reflection of monolayer graphene is negligibly small ($R \ll 1\%$), the conservation of energy directly reduces the optical scattering problem to the linear relation $T \approx 1 - A$. 

\vspace{2ex}
\noindent{\normalsize \bfseries S3. ARPES Measurements and Deformed Dirac Dispersion Curve Analyses}
\vspace{2ex}
To quantitatively analyze the electronic band structure captured by angle-resolved photoemission spectroscopy (ARPES), the energy dispersion curves were modeled based on the effective low-energy description of the graphene layers. In a full tight-binding framework for bilayer graphene, taking into account the interlayer coupling $\gamma_1$ and the potential asymmetry $\Delta_0$, the comprehensive energy dispersion spanning all four bands is thoroughly formulated in the landmark review by Castro Neto \textit{et al.} \cite{CastroNeto2009} as follows:

\begin{equation}
E_{\text{full}}(k) = \pm \sqrt{ \frac{\gamma_1^2}{2} + \Delta_0^2 + (v_F \hbar k)^2 \pm
\sqrt{ \frac{\gamma_1^4}{4} + (v_F \hbar k)^2 (\gamma_1^2 + 4\Delta_0^2) } }
\end{equation}
Although the mathematical form originates from a bilayer-like Hamiltonian, it is physically utilized here as an effective framework to phenomenologically capture the deformed Dirac-cone topology of monolayer graphene heavily perturbed by the underlying SiC substrate. 
In this expression, $v_F$ is the bare Fermi velocity, $\hbar k = \hbar \sqrt{k_x^2 + k_y^2}$ is the in-plane momentum measured relative to the $K$ ($K'$) point, and $\Delta_0$ represents the site-energy asymmetry between the sublattices induced by substrate charge transfer. 
The complex parameter $\gamma_1$, which conventionally appears in the analysis of bilayer graphene as the interlayer interaction, serves here as a critical fitting parameter to reproduce the overall profile of the distorted Dirac cone. Note that, in this framework, the site-energy asymmetry parameter $\Delta_0$ directly governs the energy gap opening at the Dirac point, defined as $E_g = 2\Delta_0$.

Consequently, by employing this full exact formulation, the model successfully captures not only the broader profile of the valence band dispersion but also the rigorous microscopic details of the ``camel-back" local minimum near the valley bottom, and the full mathematical framework provides an exact quantitative description of the strongly warped Dirac cone across the entire momentum range where our ARPES intensity is predominantly resolved.

\begin{figure}[htbp]
\centering
 \includegraphics[width=0.45\textwidth]{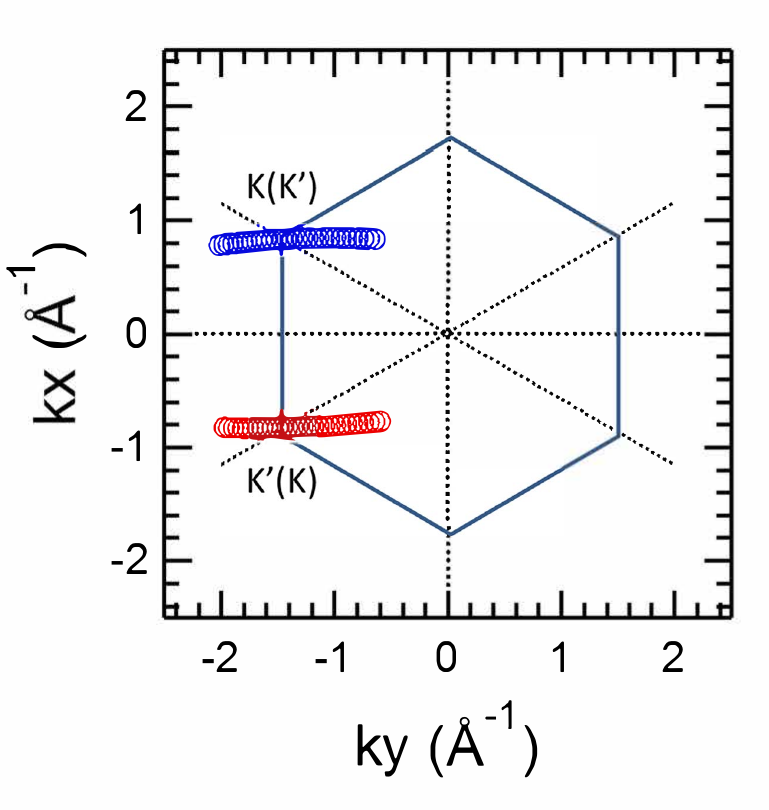}\caption{\label{fig:arpes_bz_scan}Schematic representation of the hexagonal Brillouin zone of graphene and the geometric configuration of the ARPES momentum scan. The continuous cutting line indicates the measurement pathway evaluated along the $k_y$ direction while maintaining the orthogonal momentum component $k_x$ at a constant value, directly intersecting the $K$ and $K$’ Dirac points.}
\end{figure}

The experimental trajectory for acquiring the 2D energy-momentum intensity maps is schematically illustrated in Figure~\ref{fig:arpes_bz_scan}. The photoemission spectra were recorded by systematically scanning the electron intensity along the $k_y$ direction while meticulously holding $k_x$ constant. This precise momentum slicing ensures that the cutting plane passes directly through the $K$ ($K'$) high-symmetry points at the corners of the Brillouin zone, allowing for an unskewed, high-resolution visualization of the linear Dirac-cone dispersion.

%
\vspace{2ex}
\noindent{\normalsize \bfseries S4. Alfv\'en Wave with LL-Fan Chart in Ultra High Magnetic Fields}
\vspace{2ex}

Under extreme magnetic fields, when an incident electromagnetic wave (with photon energy $\hbar\omega$) interacts with a dense, highly interacting electron-hole plasma, the system's magneto-optical response is fundamentally dominated by a collective magnetohydrodynamic excitation known as the Alfv\'en wave. The complex dielectric function $\varepsilon^*_{\pm}(\omega)$ (for right-handed ($+$) and left-handed ($-$) circularly polarized light), which dictates this collective Alfv\'en wave propagation and the overall magneto-optical spectra, is formulated as follows~\cite{bassani1975solids}:


\begin{equation}
\varepsilon^*_{\pm}(\omega) = \varepsilon_l - \frac{2\pi}{\omega} \frac{e^2}{m_0^2} \frac{\hbar}{V} \sum_{\mu,\nu} \left[ \frac{f(E_{\mu}) - f(E_{\nu})}{E_{\mu} - E_{\nu}} \frac{\left| \langle \mu | p^{\pm} | \nu \rangle \right|^2}{\hbar \omega - E_{\mu\nu} + (i\hbar / \tau_{\mu\nu})} \right]
\end{equation}      
where each physical parameter is defined by:
\begin{itemize}
   \item $\varepsilon_l$: The background (lattice) dielectric constant of the system, capturing the high-frequency polarizability of the core and substrate environment.

   \item $m_0$: The bare electron mass in vacuum.

   \item $V$: The crystal volume of the interacting electronic system.

   \item $E_{\mu}, E_{\nu}$: The eigenenergies of the single-particle Landau levels (LLs) corresponding to the initial state $|\mu\rangle$ and final state $|\nu\rangle$, calculated via our ARPES-parameterized Hamiltonian.

   \item $f(E_{\mu}), f(E_{\nu})$: The Fermi-Dirac distribution functions determining the thermal occupation probability of the respective LL states at the given Fermi energy $E_F$.

   \item $\langle \mu | p^{\pm} | \nu \rangle$: The optical dipole matrix element for the momentum operators $p^{\pm} = p_x \pm ip_y$, dictating the selection rules for circular polarization between the LL states.

   \item $E_{\mu\nu}$: The transition energy between the two Landau levels, defined as $E_{\mu\nu} = E_{\nu} - E_{\mu}$.

   \item $\tau_{\mu\nu}$: The characteristic quantum scattering lifetime (relaxation time) associated with the inter-LL optical transition, which governs the spectral broadening. In our analysis, the total spectral broadening $\Gamma_B$ incorporates both the homogeneous lifetime broadening ($\hbar/\tau_{\mu\nu}$) and spatial/structural inhomogeneous broadenings.
\end{itemize}

In this calculation, valley splitting is omitted, which is a physically well-justified approximation. Since the valley splitting induced by extreme magnetic fields ($\lesssim 10~\text{meV}$) is substantially smaller than the experimental energy resolution and the substantial spectral broadening ($\Gamma_B = 0.14~\text{eV}$), its contribution to the overall Alfvén wave absorption profile is entirely negligible.
\nocite{*}


    \end{document}